\documentclass[%
 reprint,
 amsmath,amssymb,
 aps,
]{revtex4-2}

\usepackage{graphicx}% Include figure files
\usepackage{dcolumn}% Align table columns on decimal point
\usepackage{bm}% bold math
\begin{document}

\preprint{APS/123-QED}

\title{Emergent cohesion via self-caging in maximally entangled rod packings}


\author{Yeonsu Jung}
\affiliation{%
School of Engineering and Applied Sciences, Harvard University
}%
\author{L. Mahadevan}%
 \email{lmahadev@g.harvard.edu}
\affiliation{%
School of Engineering and Applied Sciences,\\
Department of Physics,\\
Department of Organismic and Evolutionary Biology,\\
Harvard University
}%
% \altaffiliation[Also at ]{Department of Physics, Harvard University}


\begin{abstract}
Random packings of disordered rigid rods exhibit emergent cohesion, as exemplified in a nest of twigs that is self-equilibrated, free-standing structures. We analyze the geometric motif underlying this cohesion using a rod packing that maximizes the average crossing number subject to non-penetration constraints. We show that this protocol leads to self-caging: collective geometric constraints that prevent rod escape even in finite systems with free boundaries, leading to packings that remain mechanically cohesive due to a combination of purely repulsive and frictional interactions. We show that self-caging is controlled by the available free-volume in translational and rotational configuration spaces, which is minimal when $N/(Z\alpha)=1/3$ where $N$ is the number of rods, $\alpha$ is the aspect ratio, and $Z$ is the average coordination number. Our results establish a minimal geometric motif for entanglement-induced cohesion in athermal rod packings, with implications for cohesive granular matter without attractive forces.
\end{abstract}

\maketitle

%\tableofcontents

%\subsection*{Introduction}

Disordered packings with steric repulsion and dry friction exhibit collective rigidity by creating jammed structures \cite{bernal_packing_1960,ohern_random_2002,torquato_jammed_2010,liu_jamming_2010,bi_statistical_2015,zaccone_complete_2025}. 
For spheres, jamming is commonly defined as the inability to rearrange without energetic cost, a notion that typically assumes fixed boundaries or the (infinite particle) thermodynamic limit; with free boundaries, grains can always be removed from the surface. In addition, there are robust algorithms for the generation and stability of jamming \cite{makse_packing_2000,ohern_random_2002,torquato_multiplicity_2001}.  Rod packings display a distinct form of rigidity: even finite systems with free boundaries can remain cohesive because rods obstruct each other's motion through geometry, contact, and friction---a phenomenon often termed entanglement \cite{gravish_entangled_2012,karapiperis_stress_2022,patil_topological_2020,becker_active_2022,patil_ultrafast_2023,jung_entanglement_2025}. 
Despite its ubiquity, entanglement lacks both systematic protocols for its generation, and quantitative links to mechanical stability. 

Here we identify a minimal geometric motif that underlies cohesion in the random packings of rods: \emph{self-caging}. We generate maximally entangled packings by optimizing a geometric functional subject to non-penetration constraints, and show that these packings can remain cohesive under mechanical perturbations even when the interactions are purely repulsive and frictional. Unlike jammed sphere packings --where grains near a free boundary can always escape -- entangled rods packings can mutually block all lateral motions, trapping each another even with free boundaries, due to their extended nature.
% We quantify this collective trapping via free-volume measures in translational and rotational configuration space and link them directly to mechanical cohesion.

\begin{figure}[b!]
    \centering
    \includegraphics[width=1.0\linewidth]{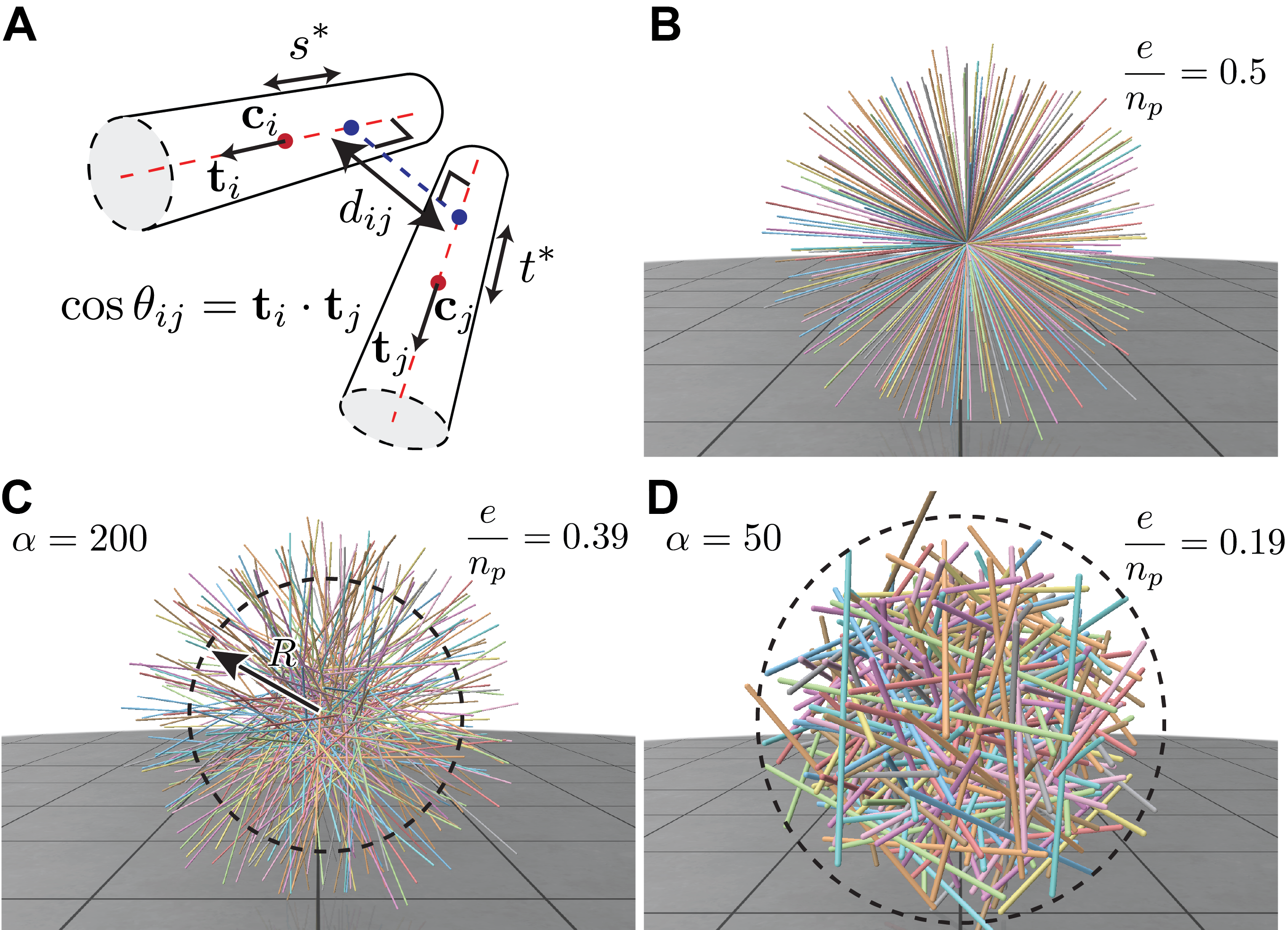}
    \caption{{\bf Maximizing entanglement for a packing with $N = 200$.}  (A) Schematic of a pair of rods. (B) Maximization of entanglement without a non-penetration constraint. Rods overlap near the global centroid. (C,D) Relaxation of the packing in (B) with an one-sided harmonic potential enforcing non-overlap, with the aspect ratios $\alpha = 200$ (C) and $\alpha = 50$ (D) The dotted circles of radius $R$ in (C) and (D) denote spheres that contain all contact points in the packing.}
    \label{fig:intro}
\end{figure}

\paragraph*{Quantification of entanglement.}

A packing of $N$ rigid spherocylindrical rods of diameter $d$ can be characterized by their generalized coordinates corresponding to the position of their centers and the orientation relative to some fixed origin $\mathbf{q}=\{ \mathbf{q}_i \}=\{(x_i,y_i,z_i,\theta_i,\phi_i) \}$ for $i=1,\dots,N$. From these generalized coordinates, we can reconstruct their centerline by $\mathbf{r}_i(s) = \mathbf{c}_i + s \mathbf{t}_i$ with $\mathbf{c}_i=(x_i,y_i,z_i)$, $\mathbf{t}_i=(\sin \theta_i \cos \phi_i, \sin \theta_i \sin \phi_i, \cos \theta_i)$, and $s\in[-l/2,l/2]$.   We fix length $l = 1$ and vary $d$ to tune their aspect ratio $\alpha = l / d$. Then the equations describing the geometry (kinematics) and dynamics of the rods can be shown to depend only on $\alpha = l/d$ (Suppl. Note 1 and 2).

We quantify entanglement using the average crossing number (ACN) which provides an upper bound for true topological measures such as various polynomial metrics \cite{buck_spectrum_2012}. For rods $i$ and $j$ with centerlines $\mathbf{r}_i(s)$ and $\mathbf{r}_j(t)$,  with unit tangent vectors $\mathbf{t}_i$ and $\mathbf{t}_j$ , we define
\begin{equation}
    \mathrm{ACN}_{ij} = \frac{1}{4\pi} 
    \int_{-l/2}^{l/2} \int_{-l/2}^{l/2}
    \frac{|(\mathbf{t}_i \times \mathbf{t}_j)\cdot(\mathbf{r}_i-\mathbf{r}_j)|}
    {|\mathbf{r}_i-\mathbf{r}_j|^3}
    \, ds \, dt,
    \label{eq:acn}
\end{equation}
which measures the average solid angle of directions for which two rods intersect in projection divided by the total solid angle ($4\pi$), since the integrand is the local solid angle of direction along $\mathbf{r}_i(s) - \mathbf{r}_j(t)$. The maximum value of pairwise ACN is 0.5 and is attained when the two rods are perpendicular and in contact with each other (Fig.~S1) consistent with intuition. In terms of an (extensive) entanglement quantity obtained for the whole network $e=\sum_{i<j}\mathrm{ACN}_{ij}$, we can define the normalized entanglement of a packing of $N$ rods as a functional of $\mathbf{q}$ \cite{becker_active_2022,jung_entanglement_2025} 
\begin{equation}
\tilde{e}(\mathbf{q})=\frac{e(\mathbf{q})}{n_p}=\frac{1}{n_p}\sum_{i<j}\mathrm{ACN}_{ij}
\label{eq:etilde}
\end{equation}
with $n_p=N(N-1)/2$.

\paragraph*{Maximizing entanglement.}

But how do we generate entangled rod packings? We do so by maximizing the entanglement measure $\tilde{e}(\mathbf{q})$ over configurations $\mathbf{q}$. Starting from initial states prepared via random sequential adsorption \cite{evans_random_1993}, we could evolve the system under the gradient flow
$\frac{d\mathbf{q}}{dt} = \nabla_{\mathbf{q}} \tilde{e}(\mathbf{q})$,
which drives the system toward a local maximizer $\mathbf{q}^* = \arg\!\max_{\mathbf{q}} \tilde{e}(\mathbf{q})$.
In practice, we employ the Fast Inertial Relaxation Engine (FIRE) \cite{bitzek_structural_2006} to efficiently reach such maximally entangled states. Since maximizing $\tilde{e}$ alone leads to unphysical overlaps between rods, we subsequently enforce non-penetration constraints using a one-sided harmonic repulsion potential activated only when $d_{ij} < d$. See Supplementary Note 4 for detailed discussion on numerical implementation.

For sufficiently large $\alpha$, the resulting relaxed configurations, $\mathbf{q}^{**}$, exhibit a dense contact-rich core  surrounding a ``thorny'' shell (Fig.~\ref{fig:intro}C). For small $\alpha$, the thorny shell structure disappear and the packing starts to look spherical (Fig.~\ref{fig:intro}D). It is worth pointing out that in sharp contrast with an isolated pair, when $\mathrm{ACN}_{ij}$ is maximized by perpendicular orientation at minimal separation, for $N > 3$ rods, the frustration due to non-penetration leads to highly non-trivial hedgehog-like structures (as in Fig.~\ref{fig:intro}B--D; see Supplementary Video 1).

% \begin{equation}
% U(x)=
% \begin{cases}
% k(x-d)^2, & x<d,\\
% 0, & x\ge d,
% \end{cases}
% \label{eq:repulsion}
% \end{equation}
% where $k$ is the effective stiffness, $x$ is the minimum inter-rod distance and $d$ is the rod diameter. 

\begin{figure}[b!]
    \centering
    \includegraphics[width=0.8\linewidth]{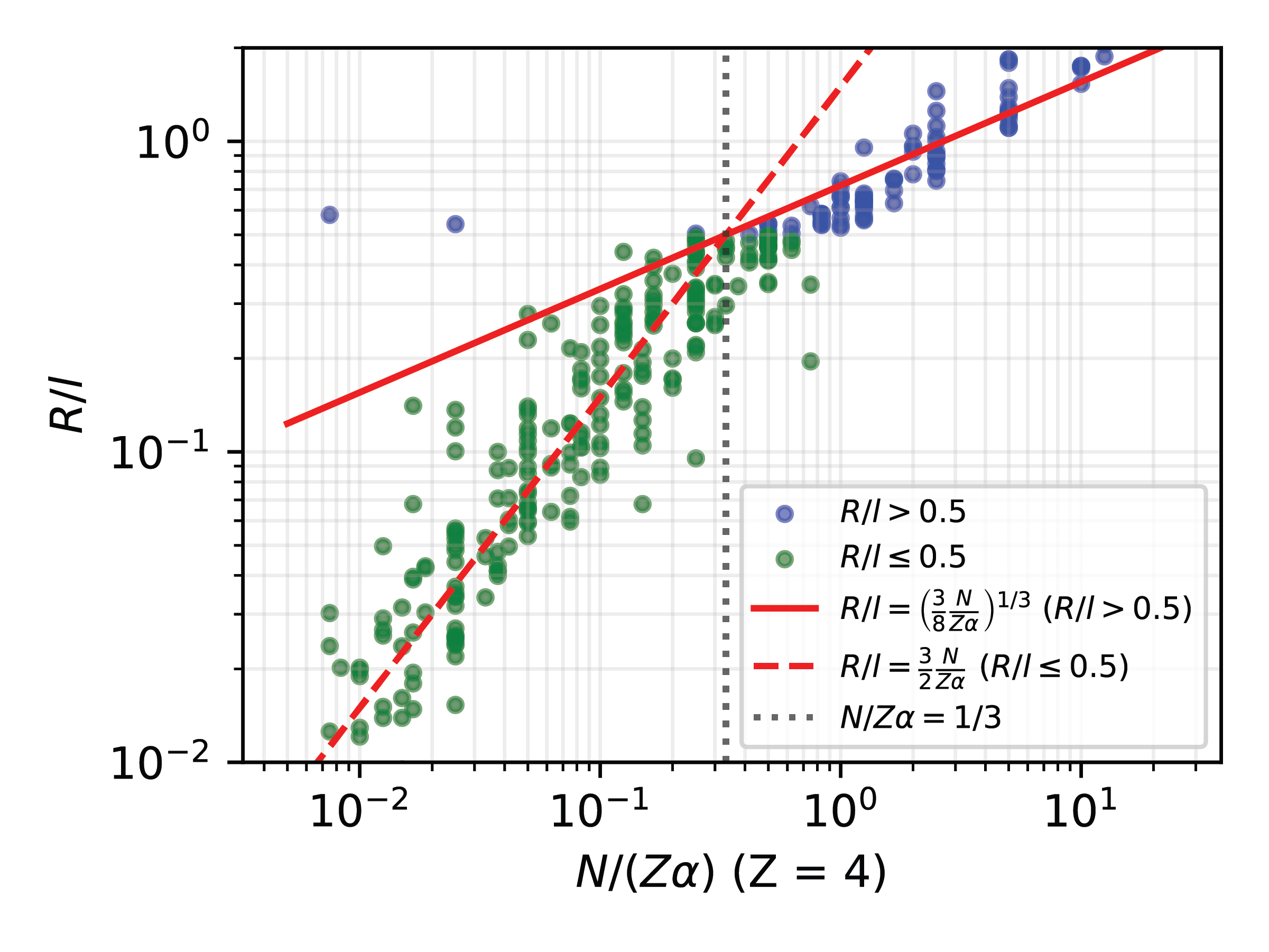}
    \caption{The normalized contact spread $R/l$ in Eq.~\ref{eq:spread} as a function of $N/(Z \alpha)$ with $Z = 4$. The theoretical predictions (red lines) are given by Eqs.~\ref{eq:one-third} and \ref{eq:linear}. }
    \label{fig:core-shell}
\end{figure}

To quantify these structures, we use the radius $R$ of the smallest sphere containing all contact points in a packing of $N$ rods with aspect ratio $\alpha$.  A packing with number density $\rho = N/V$ has an average coordination number $Z = \rho V_\mathrm{ex}$, where the excluded volume for rods of length $\zeta$ is $V_\mathrm{ex} = \pi d \zeta^2 /2$ \cite{onsager_effects_1949}. When the core radius is large compared to the rod length ($R/l > 1/2$), the length scale $\zeta$ is $l$, and subsequently $Z = NV_\mathrm{ex}/V = (3N/ 4\pi R^3) (\pi d l^2/2)$. This yields
\begin{equation}
    \frac{R}{l} = \left( \frac{3N}{2\alpha Z} \right)^{1/3}.
    \label{eq:one-third}
\end{equation}

In contrast, when $R < l/2$, the core region truncates the rods, and the effective length scale becomes $\zeta=2R$. In this regime, the same argument gives, $Z = NV_\mathrm{ex}/V = (3N/ 4\pi R^3) (2 \pi d R^2)$,
\begin{equation}
    \frac{R}{l} = \frac{3N}{2 \alpha Z},
    \label{eq:linear}.
\end{equation}

These two regimes define a crossover at $N/(Z \alpha) = 1/3$ \footnote{The solutions to $(3x/8)^{1/3} = 3x/2$ are $x = 0, -\tfrac{1}{3}, \tfrac{1}{3}$.}, separating an isotropic compact packing regime (Fig.~\ref{fig:intro}D) from the radially oriented  ``hedgehog'' regime with a dense core and an outward-pointing shell (Fig.~\ref{fig:intro}C).  To test these predictions, we compute the radius of gyration of contact points,
\begin{equation}
R^2 = \frac{1}{N_c} \sum_{i=1}^{N_c} \lvert \mathbf{p}_i - \mathbf{r}_0 \rvert^2,
\label{eq:spread}
\end{equation}
where $\mathbf{p}_i$ are contact positions \footnote{
In fully relaxed packings, $\mathbf{q}^{**}$, obtained using the one-sided repulsive potential for $d_{ij} < d$, every inter-rod distance $d_{ij}$ in $\mathbf{q}^{**}$ is strictly greater than $d$ up to a numerical tolerance of $\sim 10^{-9}$. We therefore define a contact as a pair of rods whose separation is smaller than $1.01~d$.} and $\mathbf{r}_0$ is the global centroid. 

 % The optimized packings exhibit $Z\approx 4$, a reduced coordination number relative to $Z\approx 10$ in random packings, which we attribute to the strong spatial-orientational correlations induced by entanglement maximization.

As shown in Fig.~\ref{fig:core-shell}, the measured values support the predicted scaling laws, confirming the crossover between the two regimes.

\paragraph*{Self-caging condition.}
We characterize the accessible configuration space of each rod $k$ as the connected collision-free region $D_k \subset \mathrm{SE}(3)$, and locally approximate its volume through directional free paths $r_k(\mathbf{u})$, where $\mathbf{u}$ is a unit direction in the tangent space of configuration space. For slender rods, the geometry admits two approximate symmetries: rotations about the rod axis do not affect inter-rod distances, and axial translations leave distances unchanged for sufficiently long rods. In addition, friction rapidly suppresses sliding motion. In this limit, the effective configuration space reduces to $\mathbb{R}^2 \times S^2$, corresponding to transverse translations and reorientations (see Supplementary Note 3).

\begin{figure}
    \centering
    \includegraphics[width=1\linewidth]{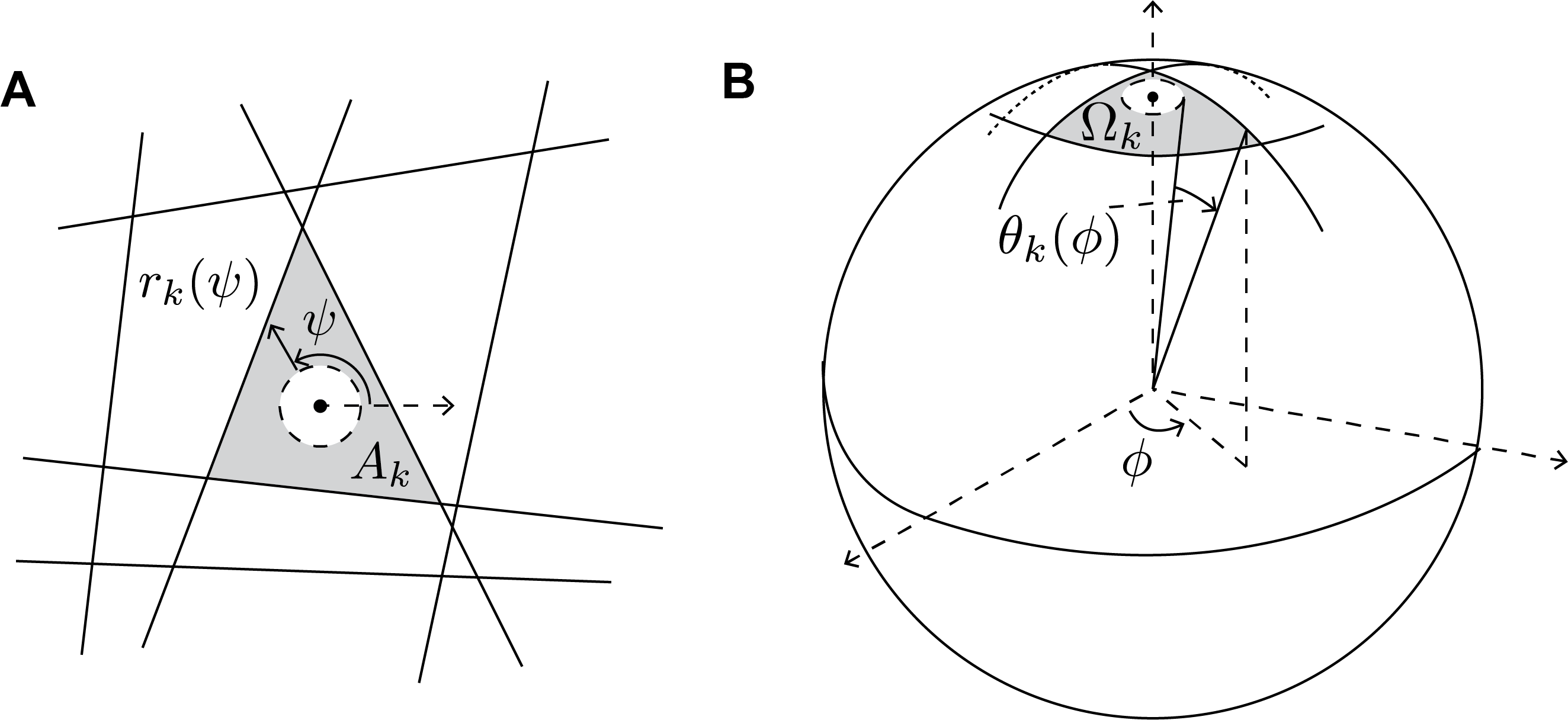}
    \caption{{\bf Self-caging property defined by free volumes in the configuration space.} (A) Translational caging diagram of a test rod $k$, obtained by projecting neighboring rods onto the plane normal to the rod axis $\mathbf{t}_k$. (B) Rotational caging diagram of a test rod $k$, obtained by projecting neighboring rods onto the unit sphere of orientations.}
    \label{fig:crossing}
\end{figure}

To characterize these motions, we define the free translational area $A_k$ and rotational solid angle $\Omega_k$ for each rod:
\begin{align}
    A_k &= \int_0^{2\pi} \frac{r_k(\psi)^2}{2} \, d\psi, \\
    \Omega_k &= \int_0^{2\pi} \left(1 - \cos\theta_k(\phi)\right) d\phi,
\end{align}
where $r_k(\psi)$ is the maximal collision-free displacement as a function of the polar angle $\psi$, and $\theta_k(\phi)$ is the maximal collision-free reorientation angle in the azimuthal direction $\phi$ (Fig.~\ref{fig:crossing}) \footnote{ For strong confinement ($\theta_k \ll 1$), the rotational measure reduces to $\Omega_k \approx \int_0^{2\pi} \theta_k(\phi)^2/2 \, d\phi$}. See Supp. Note 7 for numerical implementation.

A rod is said to be \emph{geometrically caged} if $r_k(\psi) < \infty$ for all directions $\psi$ in the plane normal to its axis. A packing is \emph{self-caged} if all rods satisfy this condition. In such configurations, rods cannot escape through lateral translation or reorientation, although axial sliding remains kinematically allowed. Mechanical stability therefore arises from the coupling between geometric confinement and friction, which arrests the remaining axial degree of freedom.

\begin{figure}[ht!]
    \centering
    \includegraphics[width=1\linewidth]{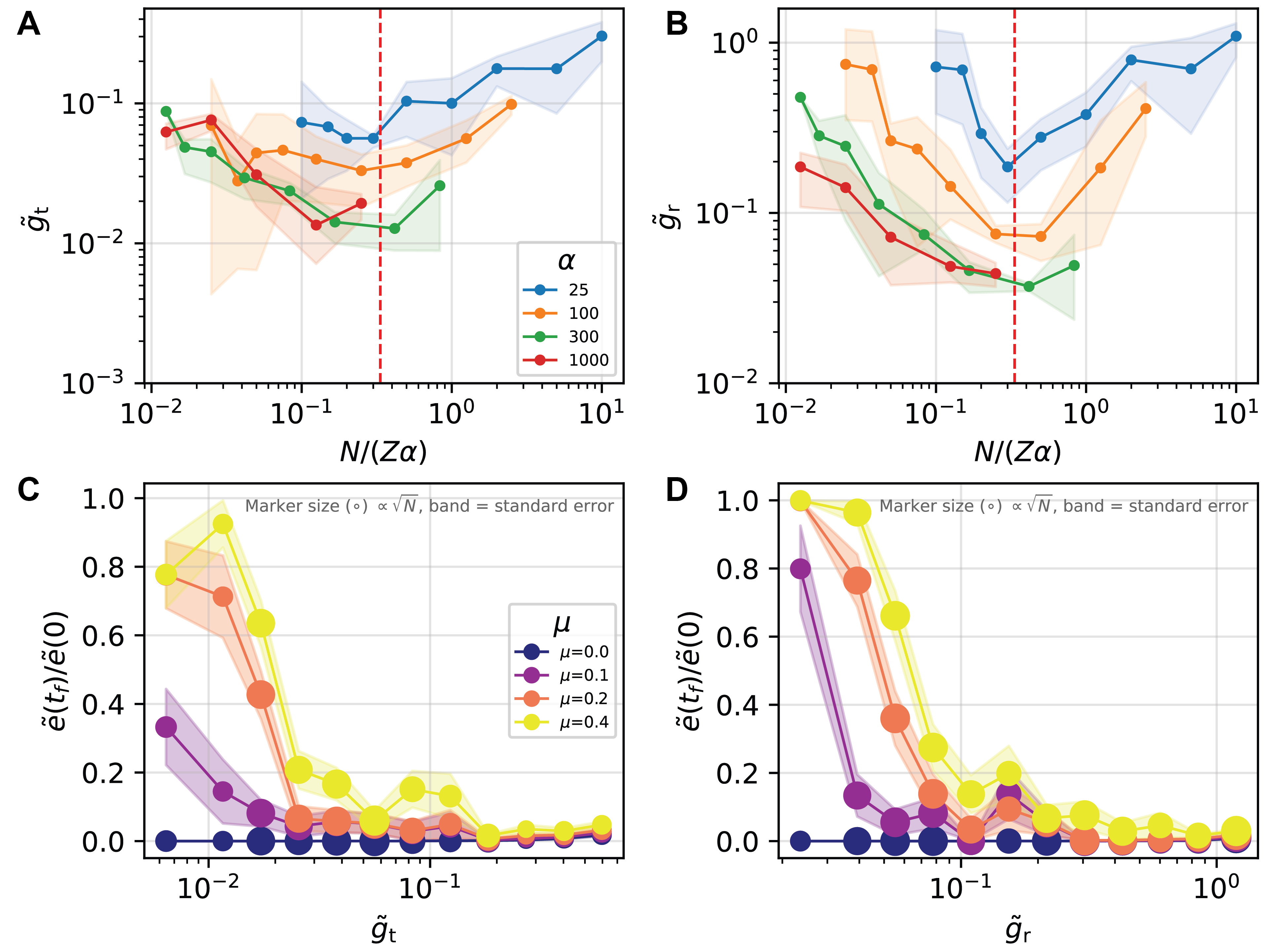}
    \caption{
    {\bf Gap-to-length ratios associated with maximum free-volume measures and their correlation with rod packing parameters entanglement retention ratio}
    (A and B) The gap-to-length ratios in Eq.~\ref{eq:gap}, $\tilde{g}_\mathrm{t}$ (translational, A) and $\tilde{g}_\mathrm{r}$ (rotational, B) as functions of $N/(Z\alpha)$, where $Z = 4$ is the reference coordination number. Lines are colored by $\alpha \in [25,100,300,1000]$ \footnote{Rigid rods with $\alpha \gg 10$ are physically unrealistic due to their vanishing bending stiffness. Here, they serve as an idealized limit to demonstrate that frictional--geometric effects, via self--caging, are sufficient to produce entanglement.}; shaded bands span the min--max range across realizations. Only realizations with finite $A^*$ are included.
    (C and D) Normalized final entanglement ratio $\tilde{e}(t_f)/\tilde{e}(0)$ as a function of $\tilde{g}_\mathrm{t}$ (C) and $\tilde{g}_\mathrm{r}$ (D), binned in equal-width intervals on a logarithmic scale. Curves are colored by friction coefficient $\mu$. Marker size scales as $\sqrt{N_\mathrm{bin}}$, where $N_\mathrm{bin}$ is the number of realizations contributing to each bin; shaded bands indicate the standard error of the bin mean. Cases with $\mu = 0$ are included to confirm that frictionless systems disentangle completely regardless of free volume.}
    \label{fig:dynamic}
\end{figure}

To quantify the tightness of self-caging, we consider the maximal free volumes
\begin{equation}
    A^* = \max_k A_k, \qquad \Omega^* = \max_k \Omega_k,
\end{equation}
which characterize the least constrained rods in the packing. We define the corresponding gap-to-length ratios
\begin{align}
\tilde{g}_\mathrm{t} = \frac{\sqrt{A^*}}{l},  ~~
\tilde{g}_\mathrm{r} &= \sqrt{ \frac{ \Omega^* }{2\pi} },
\label{eq:gap}
\end{align}
which provide dimensionless measures of translational and rotational mobility. Large values of $\tilde{g}_\mathrm{t}$ or $\tilde{g}_\mathrm{r}$ indicate weak confinement and facilitate disentanglement. Both $\tilde{g}_\mathrm{t}$ and $\tilde{g}_\mathrm{r}$ exhibit a crossover consistent with the structural transition described above. As shown in Fig.~\ref{fig:dynamic}A,B, these quantities change scaling near $N/(Z\alpha) \approx 1/3$, reflecting the transition from a core-dominated, self-caged configuration to a geometry where boundary rods are weakly constrained and mobility increases.

\paragraph*{Dynamic stability of packings.}

% ***THIS IS A REPEAT OF DISCUSSION BEFORE (6). I THINK WE CAN MOVE THE DYNAMICS SECTION ALL TOGETHER HERE, AND MOVE THE SELF-CAGING SECTION BEFORE. ****

To probe the dynamical stability of an entangled configuration $\mathbf{q}^{**}$ we subject it to a random perturbation $\mathbf{v}_0$ and evolve the system under rigid-body dynamics with contact constraints, assuming elastic collisions with restitution coefficient $\varepsilon = 1$. The configuration is constrained by non-overlap conditions,
\begin{equation}
    d_{ij} = \min_{s,t \in [-l/2,l/2]} \|\mathbf{r}_i(s) - \mathbf{r}_j(t)\| \ge d,
\end{equation}
and we define the signed distance functions
\begin{equation}
    g_{ij}(\mathbf{q}) = d_{ij} - d,
\end{equation}
so that admissible configurations satisfy $g_{ij}(\mathbf{q}) \ge 0$ for all $i<j$.
At an impact event ($g_{ij}=0$, $\dot{g}_{ij}<0$), the velocity update over a discrete time step can be written as
\begin{equation}
    \mathbf{M}(\mathbf{v}^{+} - \mathbf{v}^{-}) = \mathbf{J}^T\boldsymbol{\lambda},
    \label{eq:dynamics}
\end{equation}
where $\mathbf{v}^{-}$ and $\mathbf{v}^{+}$ are velocity before and after the collision, $\mathbf{M}$ is the mass matrix, $\mathbf{J}$ is the contact Jacobian, and $\boldsymbol{\lambda}$ are the contact impulses. The collision residual 
$\mathbf{w}=\mathbf{J}(\mathbf{v}^+ + \varepsilon \mathbf{v}^-)$ satisfies 
$\mathbf{w}\ge \mathbf{0}$, $\boldsymbol{\lambda}\ge \mathbf{0}$, and 
$\mathbf{w}^T \boldsymbol{\lambda}=0$. See Supplementary note 3 for detailed discussion including numerical implementation.

For sufficiently large initial velocities $v_0 = \lvert \mathbf{v}_0 \rvert \gg d/\Delta t$, the dense geometry of $\mathbf{q}^{**}$ induces frequent collisions over short times. At each contact, the normal impulse preserves the normal component of relative velocity, while the tangential impulse reduces the tangential component according to Coulomb friction, $\|\boldsymbol{\lambda}_t\| \le \mu \|\boldsymbol{\lambda}_n\|$. Repeated collisions therefore dissipate relative motion, driving the system towards a collective static state. The competition between sliding and collision-induced dissipation is controlled by two characteristic timescales. A rod can slide freely over a distance of order its length on a timescale $t_s \sim l/v_0$. In contrast, the mean time between successive collisions scales as $\tau_c \sim \langle g_{ij} \rangle / v_0$, where $\langle g_{ij} \rangle$ is the typical inter-rod gap. When $\langle g_{ij} \rangle / l \ll 1$, collisions occur on a much shorter timescale than sliding, suppressing axial motion and confining the dynamics to repeated contact interactions.

% This separation of scales motivates a geometric characterization of the typical gap $\langle g_{ij} \rangle$ in maximally entangled packings.

We quantify the dynamical stability of a packing by measuring the persistence of entanglement under random perturbations. To this end, we introduce an untanglement timescale $t_u$, defined as the time at which the average ACN of initially intersecting rod pairs decays to half of its initial value (see Supplementary Note 3.6). Empirically, this timescale scales as $t_u \sim d^*/v_0$, where $d^* \approx 0.32\,l$ sets a characteristic separation scale.

We measure cohesion using the normalized entanglement retention $\tilde{e}(t_f)/\tilde{e}(0)$ at a final time $t_f = 100\,t_u$. In the absence of friction, entanglement decays to zero over this timescale. In contrast, finite friction leads to sustained entanglement, with retention strongly correlated with the initial gap-to-length ratios $\tilde{g}_\mathrm{t}$ and $\tilde{g}_\mathrm{r}$ (Fig.~\ref{fig:dynamic}C,D). Packings with small $\tilde{g}_\mathrm{t}$ and $\tilde{g}_\mathrm{r}$ remain cohesive, while those with larger free volumes rapidly disentangle.

A small subset of configurations with low $N$ and large $\alpha$ exhibit weak cohesion despite small $\tilde{g}_\mathrm{t}$. These packings are marginally self-caged: translational constraints are strong, but rotational freedom remains significant, leading to eventual disentanglement. This demonstrates that translational confinement alone is insufficient for stability. The dissipation of relative motion is governed by repeated frictional collisions. When the typical gap remains approximately constant, $g \approx g_0$, the relative velocity decays algebraically in time,
\begin{equation}
    v_{\mathrm{rel}}(t) \sim \frac{g_0}{\mu} \, \frac{1}{t},
    \label{eq:powerlaw}
\end{equation}
as derived in the End Matter. Over the characteristic sliding time $t_s \sim l/v_0$, this implies $ \frac{v_{\mathrm{rel}}(t_s)}{v_0} \sim \frac{g_0}{\mu l} = \frac{\tilde{g}_0}{\mu}.$ Thus, when $\tilde{g}_0 / \mu \ll 1$, relative motion is rapidly dissipated and the packing evolves toward collective rigid-body motion.

In this regime, the remaining dynamics are dominated by global translation and rotation, with stresses supported by persistent frictional contacts (Fig.~S4). Conversely, when geometric confinement is weak, even large friction coefficients do not prevent disentanglement: although tangential velocities are locally dissipated, rods can escape through unconstrained directions (see Supplementary Videos 2–4).
These results establish that mechanical cohesion arises from a geometric--frictional coupling: geometric confinement generates sustained contacts, while friction dissipates relative motion. Neither mechanism alone is sufficient to stabilize the packing.

\paragraph*{Conclusions.}

We have shown that the dynamical stability of entangled rod packings is governed by a geometric--frictional coupling, with mechanical cohesion emerging in sufficiently well-packed slender rod assemblies. Both the free-volume measures ($A^*$ and $\Omega^*$), or corresponding gap-to-length ratios ($\tilde{g}_\mathrm{t}$ and $\tilde{g}_\mathrm{r}$) and the entanglement retention reveal a structural crossover near $N/(Z\alpha) = 1/3$, where confinement is maximized and cohesion is strongest and arises when rods mutually obstruct motion through purely repulsive interactions, generating sustained contacts that activate friction and arrest the remaining axial degrees of freedom. This defines a minimal \emph{self-caging motif} for entanglement: a dense, contact-rich core surrounded by an outward-pointing shell, whose stability is governed by collective geometric constraints rather than local packing density.

Our framework characterizes the connected collision-free region of configuration space accessible to each rod, providing a local mobility analogue of excluded-volume theories. While classical approaches integrate over global orientation distributions, the present formulation directly links geometry to collision rates and frictional dissipation. This perspective extends naturally to non-convex particles, flexible filaments, and active systems, where entanglement and self-caging govern mechanical response and emergent structure.

\vspace{10pt}

\begin{acknowledgments}
We gratefully acknowledge Mehrana Nejad, David Palmer and Sunghan Ro for extensive and insightful discussions, and partial financial support from the Harvard-NSF MRSEC-20-11754, the Simons Foundation and the Henri Seydoux Fund. 
\end{acknowledgments}

\bibliography{apssamp}% Produces the bibliography via BibTeX.

\clearpage

\section*{END MATTER}

\subsection*{Stochastic dissipation by dry friction}

The dissipation of relative motion in a contact-driven system arises from the interplay between frictional impulses and the kinematic response of the bodies at contact. During a collision, tangential relative velocities are generated by the combined translational and rotational motion of the contacting bodies, and Coulomb friction acts to oppose this motion. It is therefore useful to separate two contributions: the magnitude of the frictional impulse, controlled by the friction coefficient $\mu$, and the mechanical response of the system, determined by its inertia and contact geometry.

\begin{figure}[hb!]
    \centering
    \includegraphics[width=0.9\linewidth]{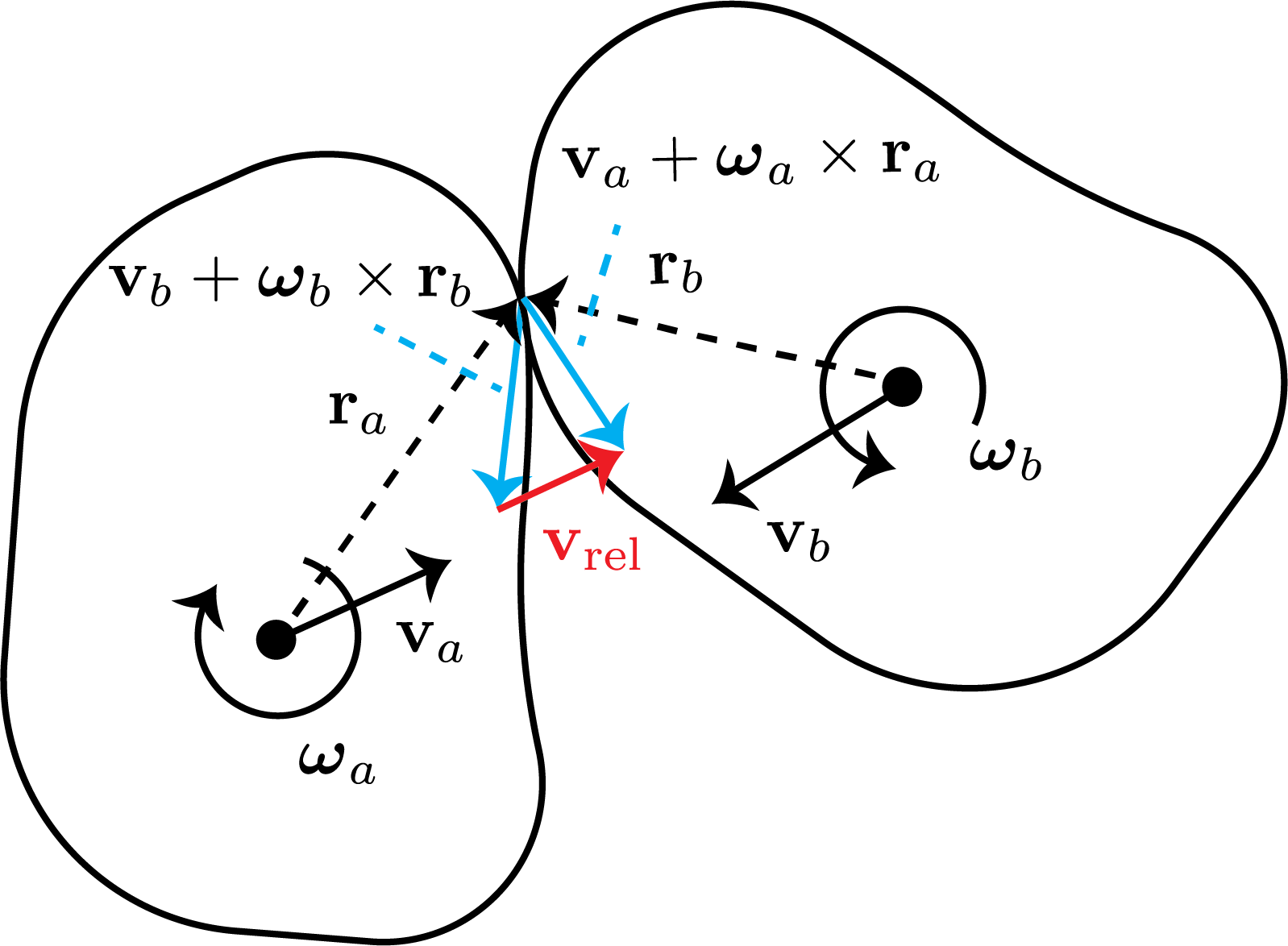}
    \caption{Schematic of two rigid bodies moving relative to each other. The relative contact velocity, $\mathbf{v}_\text{rel} = \mathbf{v}_a + \boldsymbol{\omega}_a \times \mathbf{r}_a - (\mathbf{v}_b + \boldsymbol{\omega}_b \times \mathbf{r}_b)$ depends on the contact properties (contact position and direction) as well as linear and angular velocities of the objects $a$ and $b$.}
    \label{fig:relative}
\end{figure}

This separation can be made explicit as follows. The relative contact velocity between two rigid bodies, $a$ and $b$, evaluated at the same contact point (see Fig.~\ref{fig:relative}), is
\begin{equation}
\mathbf{v}_{\mathrm{rel}}
=
\mathbf{v}_a + \boldsymbol{\omega}_a \times \mathbf{r}_a
-
\left(
\mathbf{v}_b + \boldsymbol{\omega}_b \times \mathbf{r}_b
\right),
\end{equation}
where $\mathbf{r}_a$ and $\mathbf{r}_b$ are the moment arms from the reference points of bodies $a$ and $b$ to the contact point, respectively. This relation is linear in the generalized velocity $\mathbf{v} = [
\mathbf{v}_a^T \quad
\boldsymbol{\omega}_a^T \quad
\mathbf{v}_b^T \quad
\boldsymbol{\omega}_b^T
]^T
\in \mathbb{R}^{12}.$
Thus, for this two-body contact, we can write $\mathbf{v}_{\mathrm{rel}} = \mathbf{K}\mathbf{v},$
with
\begin{equation}
    \mathbf{K}
    =
    \begin{bmatrix}
    \mathbf{I} &
    -[\mathbf{r}_a]_\times &
    -\mathbf{I} &
    [\mathbf{r}_b]_\times
    \end{bmatrix} \in \mathbb{R}^{3\times 12},
\end{equation}
where $[\mathbf{a}]_\times$ is the skew symmetric matrix associated with an arbitrary vector $\mathbf{a}$, and $[\mathbf A \quad \mathbf B]$ denotes horizontal concatenation of matrices with the same number of rows.

The operator $\mathbf{K}$ maps generalized velocities to the relative contact velocity in the lab frame. The normal and tangential components of relative contact velocity are then given by
\begin{equation}
    \mathbf{u}=
    \begin{bmatrix}
        u_n \\
        u_{t1} \\
        u_{t2} \\
    \end{bmatrix} = 
    \begin{bmatrix}
        \mathbf{n}^T \\
        \mathbf{t}_1^T \\
        \mathbf{t}_2^T \\
    \end{bmatrix}
    \mathbf{K}\mathbf{v} = \mathbf{J} \mathbf{v}
\end{equation}
where $\mathbf{n},\mathbf{t}_1$ and $\mathbf{t}_2$ denote the contact normal and two tangential unit vectors and $\mathbf{J}$  is the contact Jacobian in Eq.~\ref{eq:dynamics}. Thus, the contact Jacobian $\mathbf{J}$ is in fact the operator that maps the generalized velocity in the lab frame to the relative contact velocity in the local contact frame. In addition, given a contact impulse $\boldsymbol{\lambda}$ expressed in the local contact frame, the adjoint map $\mathbf J^T$ maps this impulse to the corresponding generalized impulse in the lab-frame generalized coordinates. Thus Eq.~\ref{eq:dynamics} is simply momentum conservation due to the local contact impulse that leads to the jump in generalized momentum $\mathbf{M}(\mathbf{v}^+ - \mathbf{v}^{-})$.

From Eq.~\ref{eq:dynamics}, we get the jump in rigid body motion, or $\mathbf{v}^{+} - \mathbf{v}^{-} = \mathbf{M}^{-1} \mathbf{J}^T \boldsymbol{\lambda}$. Applying the contact Jacobian, we get the change in relative contact velocity between a pair, or $\Delta \mathbf{u} =\mathbf{J}(\mathbf{v}^{+} - \mathbf{v}^{-})= \mathbf{J} \mathbf{M}^{-1} \mathbf{J}^T \boldsymbol{\lambda}$ where $\Delta \mathbf{u}$ is the change in the relative contact velocity. Hence, the operator $\mathbf{W} = \mathbf{J} \mathbf{M}^{-1} \mathbf{J}^T$ maps the contact impulse $\boldsymbol{\lambda}$ to the change in relative velocity $\Delta \mathbf{u} = \mathbf{J}(\mathbf{v}^{+} - \mathbf{v}^{-})$, both in the local contact frame, hence serving as ``apparent'' inverse mass operator.

Writing the local contact velocity as
\begin{equation}
\mathbf u =
\begin{bmatrix}
u_n\\
\mathbf u_t
\end{bmatrix},
\qquad
\boldsymbol{\lambda}=
\begin{bmatrix}
\lambda_n\\
\boldsymbol{\lambda}_t
\end{bmatrix},
\end{equation}

the operator $\mathbf{W}$ can be decomposed as
\begin{equation}
\Delta \mathbf u
=
\begin{bmatrix}
\Delta u_n\\
\Delta \mathbf u_t
\end{bmatrix}
=
\begin{bmatrix}
W_{nn} & W_{nt}\\
W_{tn} & W_{tt}
\end{bmatrix}
\begin{bmatrix}
\lambda_n\\
\boldsymbol{\lambda}_t
\end{bmatrix}.
\end{equation}
Thus the direct frictional contribution to the tangential velocity jump is
\begin{equation}
\Delta \mathbf u_t^{\rm fric}
=
W_{tt}\boldsymbol{\lambda}_t,
\end{equation}
while the full tangential jump also contains the coupling term
\(W_{tn}\lambda_n\).

In the sliding regime,
\begin{equation}
\boldsymbol{\lambda}_t
=
-\mu\lambda_n
\frac{\mathbf u_t}{\|\mathbf u_t\|},
\end{equation}
and the normal impulse scale is
\begin{equation}
\lambda_n\sim \frac{|u_n^-|}{W_{nn}},
\qquad
W_{nn}=\mathbf J_n\mathbf M^{-1}\mathbf J_n^T.
\end{equation}
Therefore, up to sign conventions,
\begin{equation}
\|\Delta \mathbf u_t^{\rm fric}\|
\sim
\mu
\frac{\|W_{tt}\|}{W_{nn}}
|u_n^-|.
\end{equation}
The coupling term \(W_{tn}\lambda_n\) can either increase or decrease the instantaneous tangential velocity, depending on contact geometry. In the mean-field dissipation estimate below, such normal--tangential coupling is treated as part of the fluctuating contact response factor rather than as a systematic source of damping.

For a random ensemble of contacts, we assume that the normal and tangential
components of the pre-collision relative velocity are comparable in magnitude on
average,
\begin{equation}
|u_n^-| \sim \|\mathbf u_t\| \sim v_{\rm rel}.
\end{equation}
Together with the estimate above, this implies that each sliding collision reduces the relative velocity by a fraction of order
\(\mu \Lambda^{(k)}\), where \(\Lambda^{(k)}\) is an effective
contact-dependent response factor. This motivates the mean-field recurrence
\begin{equation}
\left\langle v_{\rm rel}^{(k+1)} \right\rangle
\approx
\left(1-\mu \Lambda\right)
\left\langle v_{\rm rel}^{(k)} \right\rangle,
\end{equation}
where
\begin{equation}
\Lambda
\sim
\left\langle
\frac{\|W_{tt}^{(k)}\|}{W_{nn}^{(k)}}
\right\rangle,
\end{equation}
up to orientation, slip-direction, and normal--tangential coupling factors. Here we treat the fluctuations of \(\Lambda^{k}\) as statistically independent of \(v_{\rm rel}^{k}\) and approximately stationary as long as the entangled contact network is maintained.

Although Coulomb dissipation acts locally in the tangential direction, the randomness of contact orientations, slip directions, and contact geometries leads, at the ensemble level, to an effectively isotropic decay of the relative velocity magnitude.

The time between successive collisions depends on the velocity itself, $\tau^{k} \sim \frac{g^{k}}{v_{\mathrm{rel}}^{k}}$,
introducing a feedback mechanism: as $v_{\mathrm{rel}}^{k}$ decreases, collisions become less frequent, slowing further dissipation. In the idealized case of a constant gap, $g^{k} \approx g_0$, one has $\tau^{k} \sim g_0 / v_{\mathrm{rel}}^{k}$ (e.g., entanglement maintained). Since $v_{\mathrm{rel}}^{k} \sim (1 - \mu \Lambda)^k$ with $\Lambda=\langle \Lambda^k\rangle$, the collision times accumulate as
\begin{equation}
    t_k \sim \sum_{j=0}^{k-1} \frac{g_0}{v_{\mathrm{rel}}^{(j)}} \sim e^{\mu \Lambda k},
\end{equation}
at large $k$. Inverting this relation gives $k \sim \log t/(\mu \Lambda)$. Substituting back into the expression for the velocity yields
\begin{equation}
    v_\text{rel}(t) \sim e^{-\mu \Lambda k} \sim e^{-\log t} \sim \frac{1}{t}.
\end{equation}
Thus, the exponential decay in collision index is transformed into an algebraic decay in time through the self-consistent slowdown of collision dynamics.

This can be made explicit in a continuum description. Writing $\frac{dv_\mathrm{rel}}{dt}
    \sim
    - \frac{\mu \Lambda v_\mathrm{rel}}{\tau}
    \sim
    - \mu \Lambda \frac{v^2}{g_0},$
one obtains the effective evolution equation
\begin{equation}
    \frac{dv_\text{rel}}{dt} \sim -\mu\Lambda \frac{v_\text{rel}^2}{g_0}.
\end{equation}
Its solution is
\begin{equation}
    v_\text{rel}(t)=\frac{v_0}{1+\dfrac{\mu\Lambda v_0}{g_0}t} \sim \frac{g_0}{\mu\Lambda}\,\frac{1}{t},
\end{equation}
where the last scaling relation holds at long times, retrieving the result in Eq.~\ref{eq:powerlaw}.

This result reflects the self-delayed nature of collisional dissipation: increaseing $\mu$ increases the velocity loss per collision but simultaneously reduces the collision frequency. The combined effect produces a robust algebraic decay, with $\mu$ entering through the prefactor and setting the relaxation timescale.

In free-boundary packings, the gap $g^{k}$ evolves due to rearrangements of the contact network. As velocities decrease, collisions become less frequent and typical separations may increase, while the response factor $\Lambda^k$ may also change with geometry. These effects further slow dissipation, leading to deviations from $1/t$, including logarithmic or sub-power-law behavior, and favor sliding-dominated dynamics at long times.

\end{document}